%% file: kpnn.tex
\documentclass[12pt,fleqn,twosid,a4paper]{article}
\usepackage{a4wide}
\usepackage{epsfig}
\usepackage{slashed}
\usepackage[intlimits]{amsmath}
\usepackage[small]{caption}
\usepackage{cite}

\begin{document}

\title{
\vskip-3cm{\baselineskip14pt
\centerline{\hfill \normalsize TTP 08-09}
\centerline{\hfill\normalsize SFB/CPP-08-30}}
\vskip1.cm
\textbf{Electroweak Corrections to the Charm Quark Contribution to $K^+\to\pi^+\nu\bar{\nu}$}}

\author{
  {Joachim Brod and Martin Gorbahn}\\
  {\normalsize Institut f\"ur Theoretische Teilchenphysik,}\\
  {\normalsize  Universit\"at Karlsruhe, D-76128 Karlsruhe, Germany}\\[1em]
}

\date{}

\maketitle


\begin{abstract}
  We compute the leading-log QED, the next-to-leading-log QED-QCD, and
  the electroweak corrections to the charm quark contribution relevant
  for the rare decay \mbox{$K^+\to\pi^+\nu\bar{\nu}$}. The corresponding
  parameter $P_c(X)$ is increased by up to $2\%$ with respect to the
  pure QCD estimate to $P_c(X)=0.372\pm 0.015$ for
  $m_c(m_c)=(1.286\pm0.013)\,\textrm{GeV}$, $\alpha_s(M_Z)=0.1176\pm
  0.0020$ and $|V_{us}|=0.2255$. For the branching ratio we find
  \mbox{$B(K^+\to\pi^+\nu\bar{\nu})=(8.5\pm 0.7)\times 10^{-11}$}, where
  the quoted uncertainty is dominated by the CKM elements.
\end{abstract}


\section{Introduction}

The rare decay $K^+ \to \pi^+ \nu \bar{\nu}$ is both theoretically very
clean and highly sensitive to short-distance physics and thus plays an
outstanding role among flavour-changing neutral current processes both
in the standard model (SM) and its extensions
\cite{UT,Gino,Buras:2004uu}. Together with the process $K_L \to \pi^0
\nu \bar\nu$ it provides a critical test for the
Cabibbo-Kobayashi-Maskawa (CKM) mechanism of CP violation, while it
probes operators generated by new physics at energy scales of several
TeV \cite{D'Ambrosio:2002ex}.

In the SM the decay $K^+ \to \pi^+ \nu \bar\nu$ proceeds through
$Z$-penguin and electroweak box diagrams of $\mathcal{O}(G_F^2)$ which
exhibit a power-like GIM mechanism. This implies that non-perturbative
effects are severely suppressed and, related to this, that the
low-energy effective Hamiltonian \cite{Buchalla:1993wq,Buchalla:1998ba}
\begin{align}
\label{eq:HeffSM}
\mathcal{H}_{\text{eff}} = \frac{4G_F}{\sqrt{2}}
\frac{\alpha}{2\pi\sin^2\theta_W} \sum_{l=e,\mu ,\tau} \left(
  \lambda_c X^l(x_c) + \lambda_t X(x_t) \right) (\bar{s}_L \gamma_{\mu}
d_L) (\bar{\nu_l}_L \gamma^{\mu} {\nu_l}_L)
\end{align}
involves to an excellent approximation only a single effective operator.
Here $G_F$ is the Fermi constant, $\alpha$ the electromagnetic coupling
and $\theta_W$ the weak mixing angle.  The sum is over all lepton
flavours, $\lambda_i=V^*_{is}V_{id}$ comprises the CKM factors and $f_L$
represents left-handed fermion fields.

\begin{figure}
\centering
\includegraphics[width=13cm]{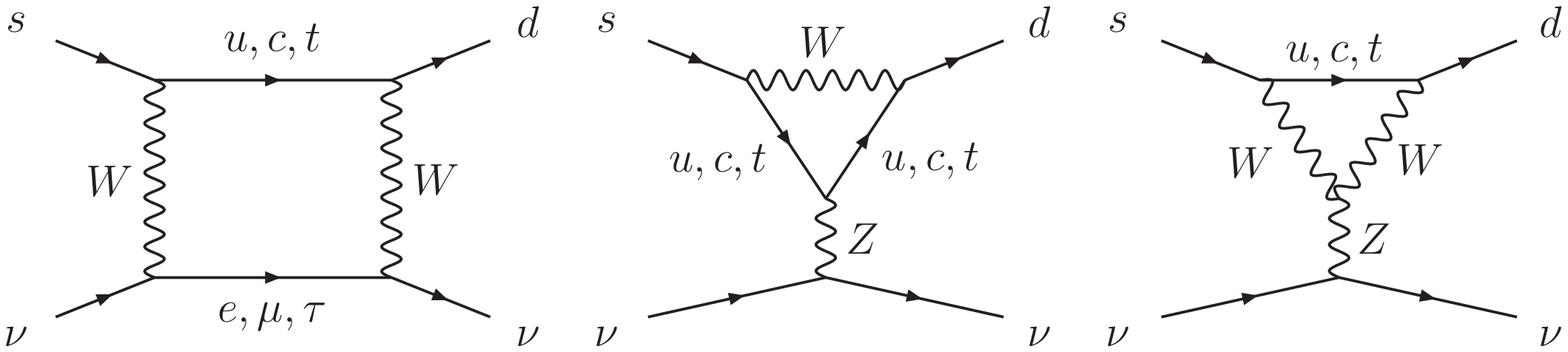}
\caption{Examples of leading-order diagrams contributing to the decay
  $K^+\to\pi^+\nu\bar\nu$ in the SM. \label{fi:kpinunu}}
\end{figure}

The function $X (x_t)$, where $x_t = m_t^2 (\mu_t)/M_W^2$ and $m_t^2
(\mu_t)$ is the top quark $\overline{\textrm{MS}}$ mass, describes the
matching contributions of internal top quarks to the operator in
Eq.~(\ref{eq:HeffSM}), where the matching is carried out at the scale
$\mu_t = \mathcal{O} (m_t)$.  Sample diagrams are shown in
Fig.~\ref{fi:kpinunu}.
The energy scales involved are of the order of the electroweak scale or
higher, while both the QCD and QED anomalous dimensions of the
corresponding operator vanish. Hence $X (x_t)$ can be calculated within
fixed-order perturbation theory.  The relevant $Z$-penguin and
electroweak box diagrams are known through next-to-leading order (NLO)
in QCD \cite{Inami:1980fz,Buchalla:1992zm, Buchalla:1998ba,
  Misiak:1999yg}. The inclusion of these $\mathcal{O} (\alpha_s)$
corrections allowed to reduce the $\pm 6 \%$ uncertainty related to the
top quark matching scale $\mu_t = \mathcal{O} (m_t)$ present in the
leading order (LO) formula down to $\pm 1 \%$. The leading term in the
large top quark mass expansion of the electroweak two-loop corrections
typically amounts to a per mil correction for the branching ratio if the
$\overline{\textrm{MS}}$ definition of $\alpha$ and $\sin^2 \theta_W$ is
used, while the uncertainty related to unknown sub-leading electroweak
contributions is conservatively estimated to be $\pm 2\%$
\cite{Buchalla:1997kz}.

The function $X^\ell (x_c)$, relevant only for $K^+ \to \pi^+ \nu \bar
\nu$,
depends on the charm quark $\overline{\textrm{MS}}$ mass through the
parameter $x_c$, conventionally defined as
\begin{equation}
  \label{eq:xcconv}
  x_c = \frac{m_c^2 (\mu_c)}{M_W^2} \, .
\end{equation}
As now both high-energy and low-energy scales, namely $\mu_W =
\mathcal{O} (M_W)$ and $\mu_c = \mathcal{O} (m_c)$ are involved, a
complete renormalisation group analysis of $X^\ell (x_c)$ is required.
In this manner, large logarithms $\ln (\mu_c^2/\mu_W^2)$ are summed to
all orders in $\alpha_s$. At LO such an analysis has been performed in
\cite{Vainshtein:1976eu}. The large scale uncertainty due to $\mu_c$ of
$\pm 26 \%$ in this result was reduced by a NLO \cite{Buchalla:1993wq,
  Buchalla:1998ba} and a subsequent NNLO calculation
\cite{Gorbahn:2004my,Buras:2005gr,Buras:2006gb} to $\pm 2.5\%$.  While
the QCD part of the calculation has reached a high level of
sophistication no QED or electroweak corrections have been included so
far. We close this gap by calculating the LO and NLO logarithmic QED
corrections as well as fixing the scheme of the input parameters in
$\sin^2 \theta_W$ and $\alpha$ by an electroweak matching calculation.
The latter point can be exemplified by noting that the charm quark
contribution is mediated by a double insertion of two dimension-six
operators. This results in a contribution of $\mathcal{O} (G_F^2)$ --
the second power of $G_F$ resides in $x_c$ -- plus electroweak
corrections. Yet the leading result of Eq.~(\ref{eq:HeffSM}) can only
approximate the electroweak corrections for a specific choice of the
renormalisation scheme for the prefactor of the charm quark
contribution, expressed as $\alpha/\sin^2 \theta_W$. While it is
expected that using $\overline{\textrm{MS}}$ parameters renormalised at
the electroweak scale would approximate the electroweak corrections best
\cite{Bobeth:2003at} only an explicit calculation can provide a definite
result. In this work we normalise all dimension-six operators to $G_F$.
Thus, we replace the parameter $x_c$ in Eq.~(\ref{eq:xcconv}) with the
unfamiliar definition
\begin{equation}
  \label{eq:xc}
  x_c = \sqrt{2} \frac{\sin^2 \theta_W}{\pi \alpha} G_F m_c^2 (\mu_c) \, ,
\end{equation}
which only at tree level equals the familiar ratio $m_c^2
(\mu_c)/M_W^2$.

The hadronic matrix element of the low-energy effective Hamiltonian can
be extracted from the well-measured $K_{l3}$ decays, including isospin
breaking and long-distance QED radiative corrections
\cite{Marciano:1996wy,Mescia:2007kn,Bijnens:2007xa}.   After summation over the three
neutrino flavours the resulting branching ratio for $K^+ \to \pi^+ \nu
\bar\nu$ can be written as\footnote{We have omitted a term which arises
  from the implicit sum over lepton flavours in $P_c$ because it amounts
  to only 0.2\% of the branching fraction. }
\cite{Buchalla:1993wq,Buchalla:1998ba,Isidori:2005xm}
\begin{multline}\label{eq:BR}
  B \left(K^+\to\pi^+\nu\bar{\nu}(\gamma)\right) \\ = \kappa_+
  (1+\Delta_{\text{EM}})
  \Bigg[\left(\frac{\text{Im}\lambda_t}{\lambda^5} X(x_t)\right)^2 +
  \left(\frac{\text{Re}\lambda_c}{\lambda} \left(P_c(X) + \delta P_{c,u}
    \right) + \frac{\text{Re}\lambda_t}{\lambda^5} X(x_t)\right)^2
  \Bigg].
\end{multline}
The parameter
\begin{equation}\label{eq:usefulP}
  P_c(X)=\frac{1}{\lambda^4}
  \left(\frac{2}{3}X^e(x_c)+\frac{1}{3}X^{\tau}(x_c)\right)
\end{equation}
describes the short-distance contribution of the charm quark, where
$\lambda= \left| V_{us} \right|$. The charm quark
contribution of dimension-eight operators at the charm quark scale
$\mu_c$ combined with long distance contributions were calculated in
Ref.~\cite{Isidori:2005xm} to be
\begin{equation}
  \label{eq:2}
  \delta P_{c,u} = 0.04 \pm 0.02 \, .
\end{equation}
The quoted error on this value can in principle be reduced with the help
of lattice QCD \cite{Isidori:2005tv}.

The remaining long distance corrections are factored out into the
following two parameters: $\kappa_+$ contains higher-order electroweak
corrections to the low energy matrix elements, and $\Delta_{\text{EM}}$
denotes long distance QED corrections. A detailed analysis of these
contributions to NLO and partially NNLO in chiral perturbation theory
has been performed by Mescia and Smith in \cite{Mescia:2007kn}, who
found the numerical values $\kappa_+ = (0.5173\pm0.0025)\times
10^{-10}(\lambda/0.225)^8$ and $\Delta_{\text{EM}} = -0.3\%$.

\section{Electroweak Corrections in the Charm Sector}\label{sec:charm}

The charm quark contribution involves several different scales and the
corresponding large logarithms have to be summed using renormalisation
group improved perturbation theory. Keeping terms to
$\mathcal{O}(\alpha_s)$ and $\mathcal{O}(\alpha/\alpha_s)$ the expansion
of the parameter $P_c(X)$ reads
\begin{equation}
P_c(X)=\frac{4\pi}{\alpha_s(\mu_c)}P_c^{(0)}(X) + P_c^{(1)}(X) +
\frac{\alpha_s(\mu_c)}{4\pi}P_c^{(2)}(X) +
\frac{4\pi\alpha}{\alpha_s^2(\mu_c)}P_c^{(e)}(X) +
\frac{\alpha}{\alpha_s(\mu_c)}P_c^{(es)}(X). 
\end{equation}
The LO term $P_c^{(0)}(X)$, the NLO term $P_c^{(1)}(X)$, and the NNLO term
$P_c^{(2)}(X)$ have been calculated in \cite{Vainshtein:1976eu}, in
\cite{Buchalla:1993wq, Buchalla:1998ba}, and in \cite{Buras:2006gb}
respectively. The main goal of the this paper is to present the
electroweak corrections $P_c^{(e)}(X)$ and $P_c^{(es)}(X)$.

The calculation is performed in two steps. First, at the scale $\mu_W
\approx M_W$ the SM is matched to an effective theory where the top
quark, the $W$ boson, and the $Z$ boson are integrated out, but the
charm quark is still a dynamical degree of freedom. Second, at the scale
$\mu_c \approx m_c$ the charm quark is integrated out and the effective
Hamiltonian in Eq.~\eqref{eq:HeffSM} is obtained.

After integrating out the particles at the electroweak scale the
effective Hamiltonian containing the dimension-six operators takes the
following form: 
\begin{equation}
  \begin{split}
    \mathcal{H}_{\text{eff}}^{\text{dim.6}} = \frac{4G_F}{\sqrt{2}}
    \bigg(
    C_W(\mu) \sum_{q=u,c}(V_{qs}Q_{3q}+V_{qd}^{\star}Q_{4q})& \\
    + \lambda_c \sum_{j=\pm} C_j(\mu)(Q_j^c-Q_j^u)& +
    \frac{1}{2} C_A(\mu) Q_A +
    \frac{1}{2} C_V(\mu) Q_V \bigg).
  \end{split}
\end{equation}
Here we kept only operators relevant for the decay
$K^+\to\pi^+\nu\bar{\nu}$. These are the semi-leptonic operators 
\begin{equation}
Q_{3q}=\sum_{l=e,\mu,\tau}(\bar{s}_L \gamma_{\mu} q_L) 
(\bar{\nu_l}_L \gamma^{\mu} l_L) \quad\text{and}\quad
Q_{4q}=\sum_{l=e,\mu,\tau}(\bar{q}_L \gamma_{\mu} d_L) 
(\bar{l}_L \gamma^{\mu} {\nu_l}_L) \, ,
\end{equation}
the current-current four-quark operators
\begin{equation}
Q_{\pm}^q = \frac{1}{2}\left((\bar{s}_L^{\alpha}\gamma_{\mu}q_L^{\alpha})   
  (\bar{q}_L^{\beta}\gamma^{\mu}d_L^{\beta}) \pm
  (\bar{s}_L^{\alpha}\gamma_{\mu}q_L^{\beta})    
  (\bar{q}_L^{\beta}\gamma^{\mu}d_L^{\alpha})\right) \, ,
\end{equation}
where $\alpha$, $\beta$ are colour indices, and the operators 
\begin{equation}
  \begin{split}
    Q_A &= 
    \sum_q \sum_{l=e,\mu ,\tau} 
    ( - I_q^3 ) (\bar{q} \gamma_{5} \gamma_{\mu}q) 
    (\bar{\nu_l}_{L} \gamma^{\mu} {\nu_l}_L) \, ,\\
    Q_V &= \sum_q \sum_{l=e,\mu ,\tau} 
    (I_q^3 - 2Q_q\sin^2\theta_W) (\bar{q} \gamma_{\mu} q)
    (\bar{\nu_l}_{L} \gamma^{\mu} {\nu_l}_L) \, ,
  \end{split}
\end{equation}
which describe the quark-neutrino interaction.  We follow
Ref.~\cite{Buras:2006gb} in the definition of the evanescent operators.
All evanescent operators relevant first at the order considered in this
work are defined as
\begin{equation}
E_{3q}^{(1)} = \sum_{l=e,\mu,\tau}(\bar{s}_L
\gamma_{\mu_1}\gamma_{\mu_2}\gamma_{\mu_3} q_L)  
(\bar{\nu_l}_L \gamma^{\mu_1}\gamma^{\mu_2}\gamma^{\mu_3} l_L) -
(16-4\epsilon) Q_{3q} \, ,
\end{equation}
i.e. the evanescent operator needed for the QED renormalisation of
$Q_{3q}$, or in an analogous way.

\begin{figure}[t]
\centering
\includegraphics[width=10cm]{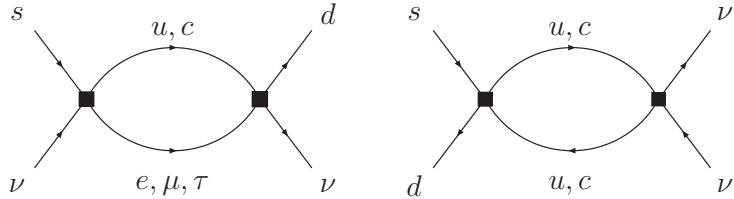}
\caption{Leading-order diagrams for the mixing of various dimension-six
  operators into $Q_{\nu}$ (see text for details).
  \label{fi:bilocalLO}}
\end{figure}

These operators mix via double insertions into the operator given in
Eq.~\eqref{eq:HeffSM}. Traditionally one distinguishes the box
contribution which comprises double insertions of the semileptonic
operators $Q_{3q}$ and $Q_{4q}$ (see Fig.~\ref{fi:bilocalLO}, left side)
and the penguin contribution which comprises double insertions of the
current-current-type operators $Q_{\pm}$ and the operators $Q_A$ and
$Q_Z$ (Fig.~\ref{fi:bilocalLO}, right side). The relevant
dimension-eight part of the effective Hamiltonian can then be written as
\begin{equation}
\mathcal{H}_{\text{eff}}^{\text{charm}} = \left(2G_F^2\lambda_c C_{\nu}^B(\mu) +
G_F^2 \lambda_c C_{\nu}^P(\mu)\right) Q_{\nu}, 
\end{equation}
where the operator $Q_{\nu}$ is defined as 
\begin{equation}\label{eq:Qnu}
Q_{\nu}= \frac{m_c^2}{g_s^2\mu^{2\epsilon}} \sum_{l=e,\mu,\tau} 
(\bar{s}_L \gamma_{\mu_1} d_L)  (\bar{\nu_l}_{L} 
\gamma^{\mu_1} {\nu_l}_{L}) \, ,
\end{equation}
while $C_{\nu}^B$ and $C_{\nu}^P$ denote the box and penguin
contribution, respectively. 

The renormalisation group analysis proceeds in several steps. The
initial conditions for the renormalisation group equations (RGE), which
govern the running of the Wilson coefficients, are calculated in
Sec.~\ref{sec:incon}. The anomalous dimensions are computed in
Sec.~\ref{sec:adm}. After integrating out the bottom and the charm
quark, the theory is matched onto the low energy effective Hamiltonian
of Eq.~(\ref{eq:HeffSM}). The relevant results are collected in
Sec.~\ref{sec:belowmuc}. In Sec.~\ref{sec:finalexp} the pieces are put
together to give the final result for $P_c(X)$.

We have computed all Feynman diagrams in this paper using \texttt{FORM}
\cite{Vermaseren:2000nd} routines and independently using
\texttt{Mathematica}. All the QCD corrections relevant to a NNLO
analysis of $P_c(X)$ are given in \cite{Buras:2006gb} and references
therein.

\subsection{Initial Conditions}\label{sec:incon}

\begin{figure}[t]
\centering
\includegraphics[width=15cm]{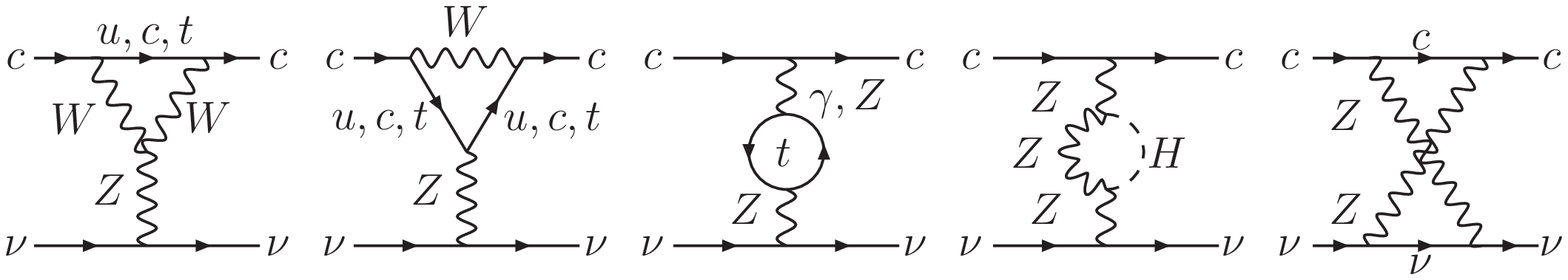}
\caption{Feynman diagrams contributing to the NLO matching for
  $C_{A}$. \label{fi:ccnuenue01}}
\end{figure}

The Wilson coefficients are found by matching the one light particle
irreducible Green's functions in the full and the effective theory at
the electroweak scale $\mu_W^2\sim M_W^2$. We use the
$\overline{\textrm{MS}}$ scheme for both theories and remark that a
finite field redefinition for the light particles ensures the correct
normalisation of the kinetic term in the effective theory. In the box
sector only $C_W$ and in the penguin sector only $C_{\pm}$ and $C_{A/V}$
receive electroweak corrections at the order considered here (see
Fig.~\ref{fi:ccnuenue01}). We expand the Wilson coefficients in powers
of the coupling constants
\begin{equation}\label{eq:Cexp}
C(\mu)=C^{(0)}(\mu)+\frac{\alpha_s(\mu)}{4\pi}C^{(1)}(\mu)+
\frac{\alpha}{\alpha_s}C^{(e)}(\mu)+
\frac{\alpha}{4\pi}C^{(es)}(\mu)
\end{equation}
and use a similar expansion for any quantity in the following, unless
explicitly stated otherwise.

We normalise the Wilson coefficients $C_W$, $C_\pm$, and $C_{A/V}$ to
the muon decay constant $G_F$ \cite{Sirlin:1981ie}. In this way most of
the radiative corrections cancel, including all terms dependent on $m_t$
and $M_H$ in the case of $C_W$ and $C_{\pm}$. All our matching
calculations have been performed in the generalised $R_\xi$ gauge for
the photon field and in the case of $C_A$ also for the $W$ and $Z$
fields as a check of our results.

At the one-loop level a neutrino-photon Green's function is generated
which contributes to $C_V$ via the equations of motion. Yet $Q_V$ does
not mix into $Q_\nu$ and the Wilson coefficient $C_V$ is not needed.

For the relevant electroweak corrections we find
\begin{equation}
  \label{eq:4}
  \begin{split}
    C_{\pm}^{(0)}(\mu_W) &= 1 \, , \\
    C_{\pm}^{(e)}(\mu_W) &= 0 \, , \\
    C_{\pm}^{(es)}(\mu_W) &= - \frac{22}{9} -
    \frac{4}{3}\ln\frac{\mu_W^2}{M_Z^2} \, ,
\end{split}
\end{equation}
in agreement with Ref.~\cite{Gambino:2001au,Gambino:2000fz},
\begin{equation}
  \label{eq:3}
  \begin{split}
    {C}_W^{(0)}(\mu_W) &= 1 \, , \\
    {C}_W^{(e)}(\mu_W) &= 0 \, , \\
    {C}_W^{(es)}(\mu_W) &= - \frac{11}{3} - 2\ln\frac{\mu_W^2}{M_Z^2}
    \, ,
\end{split}
\end{equation}
and
\begin{equation}
  \label{eq:5}
  \begin{split}
    C_A^{(0)}(\mu_W)&=1 \, , \\
    C_A^{(e)}(\mu_W)&=0 \, , \\
    C_A^{(es)}(\mu_W)&= \frac{3m_t^2}{4s_w^2 M_W^2} +
    \frac{11s_w^2-6}{4s_w^2c_w^2} - \frac{3}{4} \,
    \frac{M_W^2-c_w^2M_H^2}{(M_H^2-M_W^2)s_w^4}
    \ln{\frac{M_W^2}{M_Z^2}}\\
    &\quad + \frac{3M_H^4}{4(M_H^2-M_W^2)(M_W^2-c_w^2M_H^2)}
    \ln{\frac{M_H^2}{M_Z^2}} \, .
\end{split}
\end{equation}

\subsection{Anomalous Dimensions and RGE}\label{sec:adm}

The mixing of dimension-six into dimension-eight operators through
double insertions leads in general to inhomogeneous RGE
\cite{Herrlich:1994kh}. In the box sector they are given by
\begin{align}
  \mu\frac{d}{d\mu}C_{\nu}^B(\mu)&=\gamma_{\nu} C_{\nu}^B(\mu) +
  4\gamma_{\nu}^B C_W(\mu) C_W(\mu) \, , \label{eq:rgekpinunu1} \\
  \mu\frac{d}{d\mu}C_W(\mu)&=\gamma_W C_W(\mu)\label{eq:rgekpinunu2} \,
  ,
\end{align}
where $\gamma_W$ is the anomalous dimension of $Q_{3q}$,
$\gamma_{\nu}$ encodes the running of $Q_{\nu}$, which stems solely
from the running mass and coupling constant which in our definition
multiply the $Q_{\nu}$ operator, and $\gamma_{\nu}^B$ is the anomalous
dimension tensor of the mixing of the operators $Q_{3q}$ and $Q_{4q}$
into $Q_{\nu}$.

$\gamma_{\nu}$ is given in terms of the QCD $\beta$-function and the
anomalous dimension of the charm quark mass by
\begin{equation}
\gamma_{\nu}^{(k)}=2(\gamma_m^{(k)}-\beta_k).
\end{equation}
The explicit values are
\begin{equation}\label{eq:gammaexp}
\gamma_m^{(0)}=8 \, ,
\qquad
\gamma_m^{(e)}=\frac{8}{3} \, ,
\qquad
\gamma_m^{(es)}=\frac{32}{9} \, ,
\end{equation}
\begin{equation}\label{eq:betaqcdexp}
\beta_{0}=11-\frac{2}{3}f \, ,
\qquad
\beta_{e}=0 \, ,
\qquad
\beta_{es}=-\frac{8}{9}(f_u+\frac{f_d}{4}) \, ,
\end{equation}
where $f_u$ and $f_d$ denote the number of up- and down-type quark
flavours, and $f = f_u + f_d$. 

\begin{figure}[t]
\centering
\includegraphics[width=13cm]{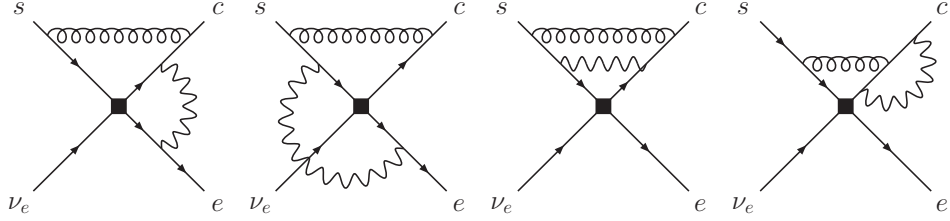}
\caption{Sample two-loop diagrams contributing to the self mixing of $Q_{3c}$. Wavy
  lines denote photons, curly lines denote gluons. \label{fi:opscenue}}  
\end{figure}

\begin{figure}[t]
\centering
\includegraphics[width=12cm]{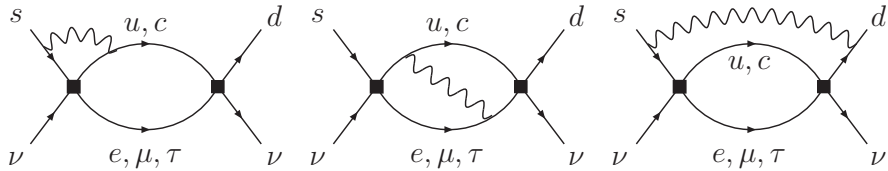}
\caption{Sample diagrams for the NLO mixing of $Q_{3q}$ and $Q_{4q}$
  into $Q_{\nu}$. \label{fi:bilocalboxNLO}}
\end{figure}

The remaining anomalous dimensions can be calculated from the pole parts of one-
and two-loop diagrams, some of which are shown in
Figs.~\ref{fi:opscenue}~and~\ref{fi:bilocalboxNLO}, using standard
methods \cite{Herrlich:1994kh,Chetyrkin:1997fm,Gambino:2003zm}. 
We find the following values: 
\begin{equation}
\gamma_{\nu}^{B(0)} = -8 \, ,
\qquad
\gamma_{\nu}^{B(e)} = 0 \, ,
\qquad
\gamma_{\nu}^{B(es)} = -\frac{316}{9} \, ,
\end{equation}
\begin{equation}
\gamma_{W}^{(0)} = 0 \, ,
\qquad
\gamma_{W}^{(e)} = -4 \, ,
\qquad
\gamma_{W}^{(es)} = 4 \, .
\end{equation}
$\gamma_{\nu}^{B(0)}$ is known for a long time (see \cite{Buchalla:1993wq}
and references therein), and $\gamma_{W}^{(e)}$ and $\gamma_{W}^{(es)}$ have already
been calculated in \cite{Sirlin:1981ie}. 

In order to solve the RGE we perform a trick
\cite{Herrlich:1996vf,Buchalla:1993wq}, so that we can use the RGE for
single insertions also in our case. To this end, we rewrite
Eq.~\eqref{eq:rgekpinunu2} as
\begin{equation}
\mu\frac{d}{d\mu}C_W^2(\mu)=2\gamma_W^T C_W^2(\mu). 
\end{equation}
Then we can combine both equations \eqref{eq:rgekpinunu1} and
\eqref{eq:rgekpinunu2} into a linear equation
\begin{equation}\label{eq:rgebox}
\mu\frac{d}{d\mu}C_B(\mu)=\gamma_B^T C_B(\mu), 
\end{equation}
where 
\begin{equation}\label{eq:Cgammabox}
C_B(\mu)=
  \begin{pmatrix} 4C_W^2(\mu)\\
                  C_{\nu}^B(\mu)
  \end{pmatrix}\quad
\text{and}\quad
\gamma_B^T=
  \begin{pmatrix} 2\gamma_W&0\\
                  \gamma_{\nu}^B&\gamma_{\nu}
  \end{pmatrix}.
\end{equation}

\begin{figure}[t]
\centering
\includegraphics[width=12cm]{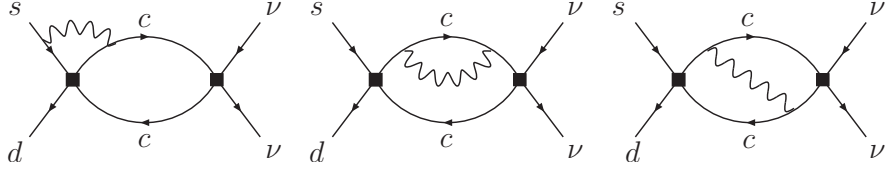}
\caption{Sample diagrams for the NLO mixing of $Q_A$ and $Q_{\pm}$ into
  $Q_{\nu}$. \label{fi:bilocalpinNLO}}
\end{figure}

The RGE for the penguin sector are given by 
\begin{align}
\mu\frac{d}{d\mu}C_{\nu}^P(\mu)&=\gamma_{\nu} C_{\nu}^P(\mu) +
4\sum_{i=\pm}\gamma_{i,\nu}^P C_i(\mu) C_A(\mu),\label{eq:rgekpinunu1pin} \\
\mu\frac{d}{d\mu}C_{\pm}(\mu)&=\gamma_{\pm}^T C_{\pm}(\mu).\label{eq:rgekpinunu2pin}
\end{align}
The anomalous dimension tensor $\gamma_{\pm,\nu}^P$ governs the mixing
of the double insertion of $Q_{\pm}$ and $Q_A$ into $Q_{\nu}$ (see
Fig.~\ref{fi:bilocalpinNLO}), while $\gamma_{\pm}$ describes the
self-mixing of $Q_{\pm}$ and was computed in \cite{Buras:1992zv}. The
anomalous dimensions read:
\begin{equation}
\gamma_{\pm,\nu}^{P(0)} = 2(1\pm 3), 
\qquad
\gamma_{\pm,\nu}^{P(e)} = 0, 
\qquad
\gamma_{\pm,\nu}^{P(es)} = \frac{52}{3}(1\pm 3) .
\end{equation}
We have defined the matrix $\gamma_{\pm}^P$ as
\begin{equation}
\gamma_{\pm,\nu}^{P(k)} = -\frac{1}{2}\gamma_{\pm,\nu}^{A(k)} -
\left(\frac{1}{2} -
  \frac{4}{3}\sin^2\theta_W\right)\gamma_{\pm,\nu}^{V(k)} \, ,
\end{equation}
with the superscripts $A$ and $V$ denoting the contributions stemming
from double insertion of $(Q_{\pm}^q,Q_A^q)$ and $(Q_{\pm}^q,Q_V^q)$,
respectively. The LO result agrees with
\cite{Buchalla:1993wq,Buras:2006gb}. The other contributions are new. 

The anomalous dimension of $Q_A$ vanishes and the RGE in the penguin
sector is the linear equation
\begin{equation}\label{eq:rgepin}
\mu\frac{d}{d\mu}C_{P}(\mu)=\gamma_{P}^T C_{P}(\mu),
\end{equation}
where 
\begin{align}\label{eq:Cgammapen}
C_{P}(\mu)&=\begin{pmatrix} 4C_+(\mu) C_A\\
                            4C_-(\mu) C_A\\
                            C_{\nu}^P(\mu)
                  \end{pmatrix},
&\gamma_{P}^T&=\begin{pmatrix} \gamma_{\pm}^T&{\begin{matrix}\gamma_{+,\nu}^P\\\gamma_{-,\nu}^P\end{matrix}}\\
                              {\begin{matrix}0&0\end{matrix}}&\gamma_{\nu}
              \end{pmatrix} \, .
\end{align}
The RGE for single insertions can be solved explicitly using the
method described in \cite{Buras:1993dy}.

\boldmath
\subsection{Below  $\mu_c$}\label{sec:belowmuc}
\unboldmath

At $\mu_c$, i.e. the scale of the charm quark mass, the charm quark is
integrated out and removed as a degree of freedom. All necessary matrix
elements are given in \cite{Buras:2006gb,Buchalla:1993wq} -- no new
contributions arise to the orders considered here. There are some new
terms stemming from the expansion of $m_c(\mu_c)$ about $m_c(m_c)$ in
these expressions, though, and we collect these results for convenience.

The matching in the box sector leads to the following matrix elements: 
\begin{equation}
r_{\tau}^{B(1)}(\mu_c) = 5 + \frac{4x_{\tau}}{1-x_{\tau}}\ln x_{\tau} +
4 \ln \frac{\mu_c^2}{m_c^2},
\end{equation}
where $x_{\tau}=m_{\tau}^2/m_c^2$ and $m_c=m_c(\mu_c)$. Neglecting the
lepton masses for the electron and muon, the above formula yields 
\begin{equation}
r_{e,\mu}^{B(1)}(\mu_c) = 5 + 4 \ln \frac{\mu_c^2}{m_c^2}. 
\end{equation}
We have defined the matrix elements for lepton flavour $l$ by 
\begin{equation}
\langle Q_l^B(\mu_c)\rangle = \frac{\alpha_s(\mu_c)}{4\pi}
r_l^{B(1)}(\mu_c) \langle Q_{\nu}\rangle^{(0)}, 
\end{equation}
where $\langle Q_l^B(\mu_c)\rangle$ denotes the double insertion of the
operators in the box sector. In the penguin sector we find
\begin{equation}
r_{\pm}^{P(1)}(\mu_c) = (1\pm 3)\left(1-\ln\frac{\mu_c^2}{m_c^2}\right),
\end{equation}
where
\begin{equation}
\langle Q_{\pm}^P(\mu_c)\rangle = \frac{\alpha_s(\mu_c)}{4\pi}
r_{\pm}^{P(1)}(\mu_c) \langle Q_{\nu}\rangle^{(0)}, 
\end{equation}
and $\langle Q_{\pm}^P(\mu_c)\rangle$ denotes the double insertion of
the operators in the penguin sector. 

\boldmath
\subsection{Final Analytic Expression for $P_c(X)$}\label{sec:finalexp}
\unboldmath

Now all that remains to do is to combine all relevant terms and compute
the box and penguin contributions to the function $X^l(x_c)$ defined in
Eq.~\eqref{eq:HeffSM}. Here we closely follow \cite{Buras:2006gb}. Let
us start with the box contribution. We expand the result as
\begin{equation}\label{eq:finalCB}
C_B^l(\mu_c) = \kappa_c\frac{x_c(m_c)}{16}
\left(\frac{4\pi}{\alpha_s(\mu_c)}C_B^{l(0)}(\mu_c) +
  \frac{4\pi\alpha}{\alpha_s(\mu_c)^2}C_B^{l(e)}(\mu_c) +
  \frac{\alpha}{\alpha_s(\mu_c)}C_B^{l(es)}(\mu_c) \right)
\end{equation}
and express the running charm quark mass $m_c(\mu_c)$ in terms of
initial condition $m_c(m_c)$, 
\begin{equation}\label{eq:xcexp}
x_c(\mu_c) = \kappa_c \left(1 +
  \frac{\alpha_s(\mu_c)}{4\pi}\xi_c^{(1)} +
  \frac{\alpha}{\alpha_s(\mu_c)}\xi_c^{(e)} +
  \frac{\alpha}{4\pi}\xi_c^{(es)} \right)x_c(m_c), 
\end{equation}
where we defined $\kappa_c=\eta_c^{(\gamma^{(0)}_m/\beta_0)}$ and
$\eta_c=\alpha_s(\mu_c)/\alpha_s(m_c)$, with the individual
contributions 
\begin{align}
\xi_c^{(1)} &= \left(\frac{\gamma^{(1)}_m}{\beta_0} -
  \frac{\gamma^{(0)}_m\beta_1}{\beta_0^2}\right) (1-\eta_c^{-1}) \, , 
\notag\\ 
\xi_c^{(e)} &= \frac{\gamma^{(e)}_m}{\beta_0} (\eta_c - 1) \, , \notag\\
\xi_c^{(es)} &= \left( \frac{\gamma^{(es)}_m}{\beta_0} -
  \frac{\beta_{es}\gamma^{(0)}_m}{\beta_0^2} -
  \frac{\beta_{1}\gamma^{(e)}_m}{\beta_0^2} \right) \ln\eta_c \notag\\ 
 &\quad  + \frac{\gamma^{(e)}_m}{\beta_0}
\left(\frac{\gamma^{(0)}_m\beta_1}{\beta_0^2} -
  \frac{\gamma^{(1)}_m}{\beta_0}\right)(1-\eta_c^{-1})(1-\eta_c).
\end{align}
We find the following expansion coefficients for $C_B^l$: 
\begin{align}
C_B^{l(0)}(\mu_c) &= C_{\nu}^{B(0)}(\mu_c),\notag\\
C_B^{l(e)}(\mu_c) &= C_{\nu}^{B(e)}(\mu_c) +
C_{\nu}^{B(0)}(\mu_c)\xi_c^{(e)} + 
4C_W^{(0)}(\mu_c)^2\rho_l^{B(e)}(\mu_c), \notag\\
C_B^{l(es)}(\mu_c) &= C_{\nu}^{B(es)}(\mu_c) +
C_{\nu}^{B(e)}(\mu_c)\xi_c^{(1)} + 
C_{\nu}^{B(1)}(\mu_c)\xi_c^{(e)} +
C_{\nu}^{B(0)}(\mu_c)\xi_c^{(es)} \notag\\ 
 &\quad + 4C_W^{(0)}(\mu_c)^2\rho_l^{B(es)}(\mu_c) +
4C_W^{(0)}(\mu_c)^2\rho_l^{B(e)}(\mu_c)\xi_c^{(1)} \notag\\ 
 &\quad + 8C_W^{(0)}(\mu_c)C_W^{(e)}(\mu_c)\rho_l^{B(1)}(\mu_c) +
4C_W^{(0)}(\mu_c)^2\rho_l^{B(1)}(\mu_c)\xi_c^{(e)}. 
\end{align}
We obtain the parameters $\rho_l^B$ by inserting the expansion
of $m_c(\mu_c)$ into the expressions for $r_l^B$ (see
Sec.~\ref{sec:belowmuc}): 
\begin{align}
\rho_{\tau}^{B(1)} &= r_{\tau}^{B(1)}(m_c) + \frac{4}{x_{\tau}-\kappa_c}
\left(\kappa_c\ln\kappa_c - \frac{x_{\tau}(1-\kappa_c)}{1-x_{\tau}}
  \ln x_{\tau}\right) \, , \notag\\
\rho_{\tau}^{B(e)} &= 0 \, , \notag\\
\rho_{\tau}^{B(es)} &= -\frac{4\kappa_c\xi_c^{(e)}
  \left[\kappa_c-x_{\tau}
    \left(1-\ln\frac{x_{\tau}}{\kappa_c}\right)\right]}{(\kappa_c -
  x_{\tau})^2}  \, .
\end{align}
The corresponding expressions for the electron and the muon, where we
can neglect the masses, are given by
\begin{align}
\rho_{e,\mu}^{B(1)} &= r_{e,\mu}^{B(1)}(m_c) - 4\ln\kappa_c \, , \notag\\
\rho_{e,\mu}^{B(e)} &= 0 \, , \notag\\
\rho_{e,\mu}^{B(es)} &= -4\xi_c^{(e)} \, .
\end{align}

The penguin contribution to the function $X^l(x_c)$ can be obtained in
the same way. Expanding the Wilson coefficients $C_P(\mu_c)$ as
\begin{equation}\label{eq:finalCP}
C_P(\mu_c) = \kappa_c\frac{x_c(m_c)}{32}
\left(\frac{4\pi}{\alpha_s(\mu_c)}C_P^{(0)}(\mu_c) +
  \frac{4\pi\alpha}{\alpha_s(\mu_c)^2}C_P^{(e)}(\mu_c) +
  \frac{\alpha}{\alpha_s(\mu_c)}C_P^{(es)}(\mu_c) \right), 
\end{equation}
we find the following contributions: 
\begin{align}
C_P^{(0)}(\mu_c) &= C_{\nu}^{P(0)}(\mu_c),\notag\\
C_P^{(e)}(\mu_c) &= C_{\nu}^{P(e)}(\mu_c) +
C_{\nu}^{P(0)}(\mu_c)\xi_c^{(e)} + 
4C_A^{(0)}(\mu_c)\sum_{i=\pm}C_i^{(0)}(\mu_c)\rho_i^{P(e)}(\mu_c), \notag\\ 
C_P^{(es)}(\mu_c) &= C_{\nu}^{P(es)}(\mu_c) +
C_{\nu}^{P(e)}(\mu_c)\xi_c^{(1)} + C_{\nu}^{P(1)}(\mu_c)\xi_c^{(e)} + 
C_{\nu}^{P(0)}(\mu_c)\xi_c^{(es)} \notag\\ 
 &\quad + 4\sum_{i=\pm}\left(\rho_i^{P(es)}(\mu_c) +
\rho_i^{P(e)}(\mu_c)\xi_c^{(1)} + \rho_i^{P(1)}(\mu_c)\xi_c^{(e)}
\right)C_i^{(0)}(\mu_c)C_A^{(0)}(\mu_c) \notag\\ 
 &\quad +
 4\sum_{i=\pm}\rho_i^{P(1)}(\mu_c)\left(C_i^{(e)}(\mu_c)C_A^{(0)}(\mu_c) +
   C_i^{(0)}(\mu_c)C_A^{(e)}(\mu_c)\right) \notag\\
 &\quad +
 4\sum_{i=\pm}\rho_i^{P(e)}(\mu_c)C_i^{(1)}(\mu_c)C_A^{(0)}(\mu_c). 
\end{align}
Again we obtain the parameters $\rho_i^P$ by inserting the expansion
of $m_c(\mu_c)$ into the expressions for $r_i^P$: 
\begin{align}
\rho_{\pm}^{P(1)} &=r_{\pm}^{P(1)}(m_c) + (1\pm 3)\ln\kappa_c, \notag\\
\rho_{\pm}^{P(e)} &= 0, \notag\\
\rho_{\pm}^{P(es)} &= (1\pm 3)\xi_c^{(e)}. 
\end{align} 

The final result for $X^l$ is then
\begin{equation}
X^l(x_c) = C_P(\mu_c) + C_B^l(\mu_c).
\end{equation}
The corresponding expressions for $C_P(\mu_c)$ and $C_B^l(\mu_c)$ can be
found in Eqs.~\eqref{eq:finalCB} and \eqref{eq:finalCP},
respectively. Eq.~\eqref{eq:usefulP} then yields the contribution to
the branching fraction. 

\section{Final Results and Numerical Discussion}

Having all necessary ingredients at hand we will discuss the numerical
implications of our results, where we use the input parameters given
in Tab.~\ref{tab:num}.
\begin{table}
\begin{tabular}{|c|l|c||c|l|c|}\hline
$M_W$&$(80.403\pm 0.029)$\,GeV&\cite{Yao:2006px}&
$\alpha_s(M_Z)$&$0.1176 \pm 0.0020$&\cite{Yao:2006px}
\\\hline
$M_Z$&$(91.1876\pm 0.0021)$\,GeV&\cite{Yao:2006px}&
$\alpha(M_Z)$&1/127.9&\cite{Yao:2006px}
\\\hline
$M_t$&$(172.6\pm 1.4)$\,GeV&\cite{Group:2008nq}&
$\sin^2\theta_W^{\overline{\text{MS}}}$&$0.23122 \pm 0.00015$&\cite{Yao:2006px}
\\\hline
$m_b(m_b)$&$(4.164\pm 0.025)$\,GeV&\cite{Kuhn:2007vp}&
$G_F$&$1.166\,37\times 10^{-5}\text{GeV}^{-2}$&\cite{Yao:2006px}
\\\hline
$m_c(m_c)$&$(1.286\pm 0.013)$\,GeV&\cite{Kuhn:2007vp}&
$\lambda$&$0.2255 \pm 0.0007$&\cite{Antonelli:2008jg}
\\\hline
$M_H$&$(155 \pm 40)$\,GeV&--&
$\left| V_{cb} \right|$ & $(4.15 \pm 0.09) \times 10^{-2}$ & \cite{Charles:2004jd}
\\\hline
$m_{\tau}$&$(1776.99^{+0.29}_{-0.26})$\,MeV&\cite{Yao:2006px}&
$\bar \rho$&$0.141^{+0.029}_{-0.017}$&\cite{Charles:2004jd}
\\\hline
&&&
$\bar \eta$&$0.343 \pm 0.016$&\cite{Charles:2004jd}
\\\hline
\end{tabular}
\caption{Input parameters used in our numerical analysis. \label{tab:num}}
\end{table}
Our numerical procedure follows closely the one of
Ref.~\cite{Buras:2006gb}. In particular we use the numerical solution of
the RGE of the program \texttt{RunDec}\cite{Chetyrkin:2000yt} to compute
$\alpha_s(\mu_c)$ from $\alpha_s(M_Z)$ and neglect all terms
proportional to $\beta_{es}$. We have checked numerically that this is
indeed justified\footnote{We thank Ulrich Haisch for providing us with
  his program for the QED running of $\alpha_s$.}.

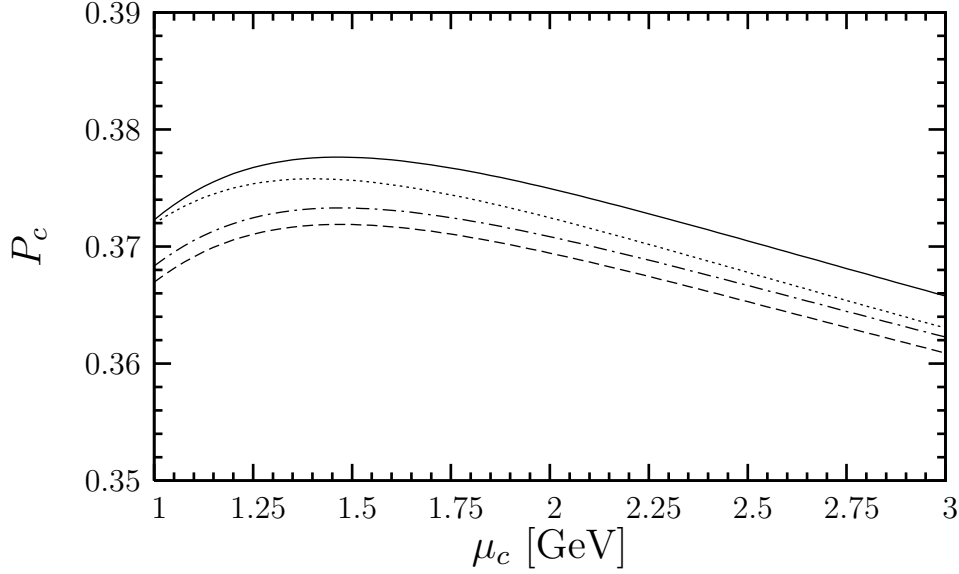
\begin{figure}
\begin{center}
\scalebox{1}{\input{pcplot.tex}}
\end{center}
\vspace{-5mm}
\caption{$P_c(X)$ as a function of $\mu_c$ at NNLO QCD (dashed dotted
  line), including LO QED (dotted line), and NLO electroweak corrections
  (solid line). The dashed line shows $P_c(X)$ at NNLO QCD where the
  definition $x_c=m_c/M_W$ is used.}
\label{fig:plots}
\end{figure}

The dependence of $P_c(X)$ on the parameter $\mu_c$ can be seen in
Fig.~\ref{fig:plots}. We use central values for all relevant input
parameters of Tab.~\ref{tab:num} and fix $\mu_b = 5\,$GeV and $\mu_W =
80\,$GeV. The dashed line shows $P_c(X)$ as a function of $\mu_c$
including the NNLO QCD corrections, as computed in \cite{Buras:2006gb}
where the parameter $x_c$ equals $m_c^2/M_W^2$. The dashed-dotted line
shows the same quantity, but using our improved definition of $x_c$, see
Eq.~\eqref{eq:xc}. We observe that this line is shifted by about 0.5\%
compared to $P_c(X)$ using the conventional definition of $x_c$. The
dotted and the solid lines show the results including LO QED and the NLO
electroweak corrections, respectively. We see that including the full
electroweak corrections, $P_c(X)$ is increased by another 1.5\% as
compared to the pure NNLO QCD result with the improved definition of
$x_c$. Also the cancellation of the scheme dependence between the LO QED
and the NLO electroweak contribution is clearly visible.

The explicit analytic expression for $P_c(X)$ including the complete
NNLO corrections is so complicated and long that we derive an
approximate formula.  Setting $\lambda = 0.2255$ and $m_t (m_t) = 163.0
\textrm{GeV}$ we derive an approximate formula for $P_c(X)$ that
summarises the dominant parametric and theoretical uncertainties due to
$m_c (m_c)$, $\alpha_s (M_Z)$, $\mu_c$, $\mu_W$, and $\mu_b$.  It reads
\begin{equation} \label{eq:masterformula}
  \begin{split}
P_c(X) &= 0.38049
\left( \frac{m_c (m_c)}{1.30 \textrm{GeV}} \right)^{0.5081}
\left( \frac{\alpha_s (M_Z)}{0.1176} \right )^{1.0192} 
\left( 1 +
  \sum_{i,j} \kappa_{ij} L_{m_c}^i L_{\alpha_s}^j
\right) \\
&\pm 0.008707 
\left( \frac{m_c (m_c)}{1.30 \textrm{GeV}} \right)^{0.5276}
\left( \frac{\alpha_s (M_Z)}{0.1176} \right )^{1.8970} 
\left( 1 +
  \sum_{i,j} \epsilon_{ij} L_{m_c}^i L_{\alpha_s}^j
\right) \, ,       
  \end{split}
\end{equation}
where 
\begin{equation} \label{eq:defLs}
L_{m_c} = \ln \left( \frac{m_c (m_c)}{1.30 \textrm{GeV}} \right) \, , 
\qquad
L_{\alpha_s} = \ln \left( \frac{\alpha_s (M_Z)}{0.1176} \right) \, ,
\end{equation}
and the sum includes the expansion coefficients $\kappa_{ij}$ and
$\epsilon_{ij}$ given in Tab. \ref{tab:kappas}. The above formula
approximates the central value of the full NNLO QCD result plus
electroweak corrections with an accuracy of $\pm 0.05 \%$ in the ranges
$1.15 \textrm{\,GeV} \le m_c (m_c) \le 1.45 \textrm{\,GeV}$, $0.114 \le
\alpha_s (M_Z) \le 0.122$, while the scale uncertainty for varying $1.0
\textrm{\,GeV} \le \mu_c \le 3.0 \textrm{\,GeV}$, $40 \textrm{\,GeV} \le
\mu_W \le 160 \textrm{\,GeV}$, and $2.5 \textrm{\,GeV} \le \mu_b \le
10.0 \textrm{\,GeV}$ is correct up to $\pm 2.3\%$ in
Eq.~(\ref{eq:masterformula}). The uncertainties due to $m_t (m_t)$, and
the different methods of computing $\alpha_s (\mu_c)$ from $\alpha_s
(M_Z)$, which are not quantified above, are all below $\pm 0.2 \%$. For
$\lambda=0.2255$ we find $P_c(X)=0.372\pm 0.015$, where $42\%$ of the
error are related to the remaining theoretical uncertainty and $58\%$ to
the uncertainties in $m_c$ and $\alpha_s$. In the future one could
utilise the correlation of $m_c$ and $\alpha_s$ in
Ref.~\cite{Kuhn:2007vp} to further reduce the parametric uncertainty.

\renewcommand{\arraystretch}{1.25}
\begin{table}[!t]
\begin{center}
\begin{tabular}{|l|l|l|l|l|}
\hline
$\kappa_{10} = 1.6624$ & 
$\kappa_{01} = -2.3537$ & 
$\kappa_{11} = -1.5862$ &
$\kappa_{20} = 1.5036$ & 
$\kappa_{02} = -4.3477$ \\
\hline 
$\epsilon_{10} = -0.3537$ & 
$\epsilon_{01} = 0.6003$ & 
$\epsilon_{11} = -4.7652$ &
$\epsilon_{20} = 1.0253$ & 
$\epsilon_{02} = 0.8866$ \\
\hline 
\end{tabular} 
\vspace{2mm}
\caption{The coefficients $\kappa_{ij}$ and $\epsilon_{ij}$ arising in
  the approximate formula for $P_c(X)$.}
\label{tab:kappas}
\end{center}
\end{table}

Finally we provide an updated number for the branching ratio: 
\begin{equation}\label{eq:BRnum}
  B\left(K^+\to\pi^+\nu\bar{\nu}(\gamma)\right) =
  (8.51^{\, +0.57}_{\, -0.62} \pm 0.20 \pm 0.36)\times 10^{-11}. 
\end{equation}
The first error stems from the uncertainties in the CKM parameters. The
second error is related to the uncertainties in $m_c$, $m_t$, and
$\alpha_s$, where all three quantities contribute in equal shares. The
dependence on $M_H$ is completely negligible (below one per mil). The
last error quantifies the remaining theoretical uncertainty. Here the
main contributions stem from the uncertainty in $\delta P_{c,u}$ and
$X_t$, where we used an error of $2\%$. In detail, the contributions to
the theory error are ($\kappa_{\nu}^+: 6\%$, $X_t: 38\%$, $P_c: 17\%$,
$\delta P_{c,u}: 39\%$), respectively. All errors have been added in
quadrature.

\section{Conclusion}

In this paper we have calculated the $\mathcal{O}(\alpha)$ and
$\mathcal{O}(\alpha \alpha_s)$ anomalous dimensions and the electroweak
matching corrections of the charm quark contribution relevant for the
rare decay $K^+ \to \pi^+ \nu \bar \nu$. The parametric dependence of
the relevant parameter $P_c(X)$ plus its theoretical uncertainty is
summarised in an approximate but very accurate formula.

$P_c(X)$ is increased by up to 2\% as compared to the previously known
results \cite{Buras:2006gb}. This change is of the same order of
magnitude as the remaining scale uncertainties after the NNLO QCD
calculation. Together with the recently achieved very precise
determination of the hadronic matrix elements \cite{Mescia:2007kn},
further improvements on the long-distance contribution of the charm
quark \cite{Isidori:2005tv}, and the complete electroweak matching
corrections for the top quark contribution \cite{BG} the theoretical
prediction of the branching ratio $B(K^+\to\pi^+\nu\bar{\nu})$ will
reach an exceptional degree of precision, with the uncertainties mainly
due to the CKM parameters.

The latter errors will be reduced in the coming years by the $B$-physics
experiments and a precise measurement of the branching ratio
$B(K^+\to\pi^+\nu\bar{\nu})$ will provide a unique test of the flavour
sector of the SM and its extensions.

\subsection*{Acknowledgements}

We would like to thank Andrzej Buras, Ulrich Haisch, and Ulrich Nierste
for their careful reading of the manuscript. We are especially grateful to
St\'ephanie Trine and Christopher Smith for interesting discussions and
comments on the manuscript. The work of JB is supported by the EU
Marie-Curie grant MIRG--CT--2005--029152 and by the DFG--funded
``Graduiertenkolleg Hochenergiephysik und Teilchenastrophysik'' at the
University of Karls\-ruhe.

\newpage

\end{document}

%% file: pcplot.tex
\begingroup%
  \makeatletter%
  \newcommand{\GNUPLOTspecial}{%
    \@sanitize\catcode`\%=14\relax\special}%
  \setlength{\unitlength}{0.1bp}%
\begin{picture}(3600,2160)(0,0)%
\special{psfile=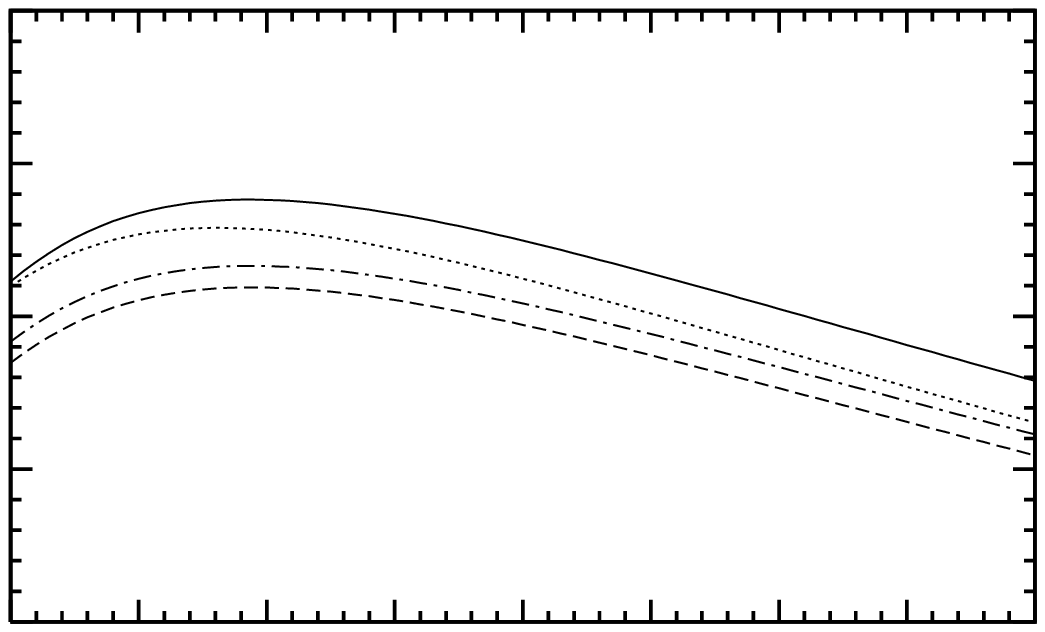 llx=0 lly=0 urx=360 ury=216 rwi=3600}
\put(1975,50){\makebox(0,0){\scalebox{1.3}{$\mu_c$ [GeV]}}}%
\put(100,1180){%
\special{ps: gsave currentpoint currentpoint translate
270 rotate neg exch neg exch translate}%
\makebox(0,0)[b]{\shortstack{\scalebox{1.3}{$P_c$}}}%
\special{ps: currentpoint grestore moveto}%
}%
\put(3450,200){\makebox(0,0){ 3}}%
\put(3081,200){\makebox(0,0){ 2.75}}%
\put(2713,200){\makebox(0,0){ 2.5}}%
\put(2344,200){\makebox(0,0){ 2.25}}%
\put(1975,200){\makebox(0,0){ 2}}%
\put(1606,200){\makebox(0,0){ 1.75}}%
\put(1238,200){\makebox(0,0){ 1.5}}%
\put(869,200){\makebox(0,0){ 1.25}}%
\put(500,200){\makebox(0,0){ 1}}%
\put(450,2060){\makebox(0,0)[r]{ 0.39}}%
\put(450,1620){\makebox(0,0)[r]{ 0.38}}%
\put(450,1180){\makebox(0,0)[r]{ 0.37}}%
\put(450,740){\makebox(0,0)[r]{ 0.36}}%
\put(450,300){\makebox(0,0)[r]{ 0.35}}%
\end{picture}%
\endgroup
 